\documentclass[preprint,aip,jcp]{revtex4-1}

\usepackage[paper=a4paper,dvips,top=1.5cm,left=1.5cm,right=1.5cm,
    foot=1cm,bottom=1.5cm]{geometry}

\usepackage{times}
\usepackage{graphicx}
\usepackage[fleqn]{amsmath}
\usepackage{amsfonts}
\usepackage{amssymb}
\usepackage{amsthm}
\usepackage{amsopn}
\usepackage{xspace}
\usepackage{array}
\usepackage{epsfig}
\usepackage{booktabs}
\usepackage{multirow}
\usepackage{subfig}
\usepackage{xcolor}
\usepackage{url}

\newcommand{\angstrom}{\mbox{\normalfont\AA}}

\begin{document}

\title{Computational Prediction of Muon Stopping Sites Using Ab Initio Random Structure Searching (AIRSS)}
\author{Leandro Liborio}
\author{Simone Sturniolo}
\author{Dominik Jochym}
\affiliation{Scientific Computing Department, Rutherford Appleton Laboratory, STFC, Didcot, United Kingdom, OX11 0DE}
\date{\today}

\begin{abstract}
The stopping site of the muon in a muon-spin relaxation experiment ($\mu$+SR) is in general unknown. There are some techniques that can be used to guess the muon stopping site, but they often rely on approximations and are not generally applicable to all cases.  In this work, we propose a purely theoretical method to predict muon stopping sites in crystalline materials from first principles.  The method is based on a combination of ab initio calculations, random structure searching and machine learning, and it has successfully predicted the Mu$_{T}$ and Mu$_{BC}$  stopping sites of muonium in Si, Diamond and Ge, as well as the muonium stopping site in LiF, without any recourse to experimental results.  The method makes use of Soprano, a Python library developed to aid ab-initio computational crystallography, that was publicly released and contains all the software tools necessary to reproduce our analysis.
\end{abstract}

\maketitle
 
\section{Introduction} 

In a muon-spin relaxation experiment ($\mu$+SR) spin-polarized positive muons are implanted in a sample to probe its local static and dynamic magnetic properties. $\mu$+SR is a sensitive probe of magnetism, but one of its limitations is not knowing the site of implantation of the muon, which limits its use for measuring magnetic moments or for comparing different magnetic structures. \\
	
There are some techniques that can determine the muon stopping sites, but these techniques are limited to certain specific cases.  For instance, in some materials, the determination of the muon stopping site was possible by using accurate experimental studies of the muon frequency shift in an applied magnetic field \cite{Renzi:1984aa}.  In particular, the so-called channelling and blocking techniques have produced some experimental information on the location of the muon site in semiconductor materials \cite{Patterson:1988aa}.\\

Nonetheless, the number of examples where the muon site can be determined by experimental means alone is limited, and in materials such as Fe$_{3}$O$_{4}$a combination of experiments and calculations has been used to determine the muon stopping sites \cite{Renzi2}.  In these examples, theoretical calculations are a cheap way of testing potential muon stopping sites.  For instance, the muon is placed in a particular site and one of the hyperfine coupling constants (HFCC) of muonium in that site is calculated and compared with the experimental results to check the validity of the muon stopping site.  However, in most of these examples there are a few starting guess sites for the muon, which limits the number of test calculations. This limiting of the potential stopping sites is not possible in most systems and, therefore, it is not always possible to determine the muon stopping site using these combined techniques. \\	

Whenever the candidate muon sites cannot be assigned by an educated guess that uses experimental data, we need to explore all the possible interstitial sites using a theoretical method. This can be a radically different process depending on whether the muon remains in its charged state ($\mu^+$) or captures an electron forming the pseudo atom muonium (Mu). One of the currently most popular methods for predicting the stopping site is called the Unperturbed Electrostatic Potential (UEP) method.  This method relies on the analysis of the electrostatic potential of the host material, which is obtained from DFT simulations. The UEP method is not fully reliable.  For instance, it has predicted the stopping sites of charged muons in materials such as RFeAsO (R=La, Ce, Pr, and Sm) \cite{PhysRevB.80.094524} and LaCoPO \cite{PhysRevB.87.064401}, but could not predict the stopping sites of neither $\mu^+$ nor Mu in fluorides \cite{moller2013quantum, PhysRevB.87.115148}. In the case of muonium, it has been suggested that the screening provided by the electron makes the muon less sensitive to the electrostatic potential of the host \cite{PhysRevB.60.13534, PhysRevB.55.6927, moller2013quantum} .  With regards to charged muons, it has been proposed that the inability  to account for the muon-host interactions is what decreases the efficiency of the UEP method in predicting the muon stopping sites \cite{moller2013quantum, PhysRevB.87.115148}. Hence, we believe that an alternative method is needed, and we propose one that combines machine learning methods with a computational technique known as Ab Initio Random Structure Searching (AIRSS) \cite{PhysRevB.78.184102, PhysRevB.80.144112}.\\

The AIRSS approach makes use of random structure generation to perform an unbiased search of stable crystal structures for a given stoichiometry \cite{pickard2011ab}. Constraints encoding whatever knowledge we have on the system, like symmetries, can be included. Our aim is to automate as much as possible the process of extracting information about not only the absolute minimum of the system, but also the relative ones. In this work, we focus on the paramagnetic states formed by muons in semiconductors. In particular, we revisit the case of muons in pure Si, Ge, Diamond and LiF \cite{moller2013quantum}, and use a combination of computational methods to retrieve the known muon stopping sites for these systems.  We first follow the established AIRSS methodology by generating a number of muoniated random structures and running ab-initio geometry optimization calculations on them. Then we apply our new machine learning methods to analyse the resulting structures and use them to predict the muon stopping sites.\\ 

\section{Methodology}

\subsection{Structure Generation}

Ab-Initio Random Structure Search, or AIRSS \cite{airss:2011}, has been demonstrated as a successful approach to many complex problems in crystalline structure detection; by comparison with most of them, the problem of finding the optimal stopping site of muonium in a crystal is much simpler, since it only involves a configuration space defined by three degrees of freedom, rather than dozens or hundreds. For this reason we decided to apply this method as a starting point for our calculations.\newline
The AIRSS approach consists of generating random initial configurations and then using CASTEP (or another equivalent DFT code) to perform a geometry relaxation on each of them. The rationale for this approach is that studies of potential energy surfaces for crystals shows that the attraction basins of the various local minima tend to have a greater hypervolume the lower the minimum's energy is \cite{doye1998thermodynamics,doye2005characterizing,massen2007power}. In other words, by sampling in a uniform random way the entire configuration space, it is statistically more likely that the great majority of starting configurations will relax to a relatively stable energy minimum. However, given the expense of performing these calculations, it is sensible to trim down some of the most unreasonable starting configurations to avoid wasting time. This was done in two steps:

\begin{itemize}
\item First, all configurations in which the muonium collided with an existing atom were eliminated. This is part of the standard AIRSS methodology;
\item Second, all configurations were examined, and when a pair was found with starting positions closer than a given limit, one of the two was eliminated. This was done to avoid running redundant calculations on configurations that would likely slide into the same local minimum, by approximating what in computer graphics is known as a "Poisson sphere" distribution \cite{lagae2006poisson}. This was done with an in house Python script.
\end{itemize}

The radii used to apply both these filtering processes can be found in table \ref{tab:radii}.  \textcolor{red}{The random nature of the process used to generate the configurations prevents us from estimating exactly the number of initial structures that will be generated with given input parameters.  Changing the values of the $r_{A-Mu}^{min}$ and $r_{Mu-Mu}^{min}$ distances, within meaningful physical limits, will alter the number of initial structures.  However, after relaxing the configurations, any redundant configurations  would likely slide into the same local minimum, without affecting the final results. In general therefore it is necessary to use distances that are small compared to what we expect the size of the attraction basis for the potential minima to be. For $r_{A-Mu}^{min}$ this is somewhat more intuitive, as we know the typical bonding distances between atoms and can pick a value that is small compared to that, assuming that any shorter distance would surely result in a collision and a repulsive interaction. There is no hard and fast rule for $r_{Mu-Mu}^{min}$, but the values used here were based just on a physical intuition of how smooth we may expect the potential landscape to be for a unit cell of this size. In general, a conservative choice would lean towards smaller values, which allow for a greater likelihood of finding potential minima with small basins; the trade-off is of course that this choice would correspond to more structures generated and thus more expensive calculations.}\newline
When generating the structures, particular care was taken to take into account the effects of periodicity, so that every time a distance between a pair of atoms was calculated, this was reduced to the distance between the two closest periodic copies of those atoms, avoiding artefacts due to the particular choice of unit cell representation.

\begin{table} 
\begin{tabular}{ | c | c | c | }
\hline
  Atom & $r_{A-Mu}^{min}$ (\AA)& $r_{Mu-Mu}^{min}$ (\AA)\\
\hline
C & 0.7 & 0.7 \\
Si & 1.1 & 0.7 \\
Ge & 1.0 & 0.7 \\
Li & 1.8 & 0.8 \\
F & 0.7 & 0.8 \\
\hline  
\end{tabular}
\caption{Distances (in Angstroms) used to discard AIRSS generated starting configurations for diamond, silicon, germanium and lithium fluoride. $r_{A-Mu}^{min}$ refers to the minimum distance between a muonium and an atom of the host crystal, while $r_{Mu-Mu}^{min}$ refers to the minimum distance between two muonium starting positions.} 
\label{tab:radii}
\end{table}

\subsection{Ab Initio Computational Details }  \label{computational-details}

We performed  Density functional theory  (DFT)  calculations with the CASTEP~\cite{Clark:2005jh} code within the generalized gradient approximation (GGA)~\cite{Perdew:1997gs} and using ultrasoft pseudopotentials~\cite{Lin:1993fz}.  These calculations were performed in  Si, Ge, Diamond and LiF  2$\times$2$\times$2 supercells based on the materials' conventional cubic unit cells and which contained one muonium each that had been placed in a random position.  These supercells were big enough to prevent interactions between the muons placed in the periodic images of the supercell created by the DFT calculations.   All calculations were spin polarised, with the initial spin placed on the muonium atom, and all were performed with a wavefunction cutoff of 450~eV for Si, 550~eV for Ge, 600~eV for Diamond and 700~eV for LiF.  We used Monkhorst-Pack grids~\cite{PhysRevB.13.5188} of 2x2x2, for sampling the reciprocal space of Si, Ge and Diamond, and of 3x3x3 for sampling the reciprocal space of LiF.  All the ions were allowed to relax until the  total energy change and forces in all ions had fallen below convergence thresholds of 1$\times$10$^{-8}$~eV and 1$\times$10$^{-4}$~$\frac{\normalfont{eV}}{\angstrom}$ respectively.  The relaxations were performed constraining the lattice parameters of the cells to their experimental values of Si=5.430$\angstrom$, Ge=5.652$\angstrom$, Diamond=3.567$\angstrom$ and LiF=4.02$\angstrom$.

\subsection{Cluster Analysis}\label{cluster-analysis}

A common way of analysing AIRSS results is to simply classify them based on their energy. However, in this work we pioneer a more advanced method based on machine learning techniques, by which we try to classify the output structures combining energy and geometric parameters, with the aim of extracting more information that otherwise might go unnoticed. This is motivated by our interest in all potential stopping sites, not just the lowest energy one, which requires us to be able to recognise different configurations in the higher energy range. Intuitively, we expect that if suitable parameters are chosen to describe their key properties, all final configurations that represent random fluctuations around a specific stopping site will look far more similar to each other than to those around a different site. This intuition, that would make it easy for a human eye to recognize the sites by looking at a 3D representation of the various structures, is what we try to automate in a way that might allow us to sift through dozens or hundreds of candidates far more efficiently and quickly than any human would.\newline
The technique used here is implemented in the Python library Soprano. Soprano has been developed with funding from the CCP for NMR Crystallography, is licensed under the GNU LGPL and can be downloaded for free \cite{soprano}. The main purpose of Soprano is to provide the users with a set of tools to create, manipulate and analyse large amounts of chemical structures, building upon the well known Atomic Simulation Environment (ASE) \cite{ase-paper} library. In this specific case we make use of the `phylogenetic' analysis tool, which characterises each structure with an array of user-defined properties and then clusters them by similarity, using the algorithms implemented in the \texttt{cluster} package of the Scipy library \cite{oliphant2007python}. These algorithms constitute the simplest type of unsupervised learning for pattern classification. Their purpose is to split an ensemble of points in an N-dimensional space (here the dimensionality is controlled by the number of parameters we choose to use for classification) based on their distance; different metrics can be used sometimes, but in this work we stick to the traditional Euclidean one, $\mathbf{r}^2 = \sum_i x_i^2$. There are two principal clustering algorithms implemented in Scipy:

\begin{itemize}
\item the \textit{hierarchical} clustering method forms clusters by iteratively clumping together points. It will look for the closest point-point, cluster-cluster, or cluster-point pair, join them together, then repeat the procedure until the shortest distance exceeds a user-defined 'tolerance' parameter. This tolerance is usually referred to as $t$ and is normalised to the maximum distance in the system, so that for $t=1$ only one cluster will exist, and for $t=0$ all points will constitute their own clusters. For point-cluster distances, the cluster is represented by its closest point to the other point (in the case of two clusters, all possible combinations are considered and the shortest one is picked). When using this method it is possible to build a dendrogram showing how the various points join up in clusters as $t$ increases; the lowest the value of $t$ at which a cluster is formed, the more similar its member elements will be;
\item the \textit{k-means} clustering method instead takes as an initial parameter a guess for the expected number of clusters $k$. It then proceeds to create $k$ centres for the clusters, and to attribute the points each to a cluster based on which centre is the closest. After that, the centres are moved to minimise the overall root sum square of distances, and the assignment is repeated. The procedure continues self-consistently until the clusters' composition stops changing and the RSS is minimised.
\end{itemize}

Our method has been to use both these algorithms in sequence. First, the hierarchical method is employed, and a dendrogram is plotted to have a bird's eye view of the structure of the set. If strongly distinct clusters are present, as it should be the case if the AIRSS run did indeed produce examples of multiple stopping sites, then this should be easily visible. We can then use the dendrogram to estimate how many clusters we could expect, and input that parameter in the $k$-means algorithm. If the classification is meaningful, there should be reasonable agreement between the size and composition of the clusters found with either method.\newline
Information on how the arrays identifying each system are built in Soprano is provided in Appendix \ref{app:phylogen}. Here we focus instead on the parameters chosen as physically significant for this specific case study. Our choice fell on two of the `genes' implemented by default in Soprano. The first was simply the energy. The second was a collection of Steinhardt bond order parameters, \cite{steinhardt1983}. These are known rotationally invariant functions which describe the local environment of a given atom with the use of spherical harmonics; they can be considered a sort of power spectrum for angular frequencies instead of spatial ones. Traditionally, these are used for probing local order in disordered systems such as liquids or glasses, and that makes them especially sensitive to the shape of the site in which a muon sits, even accounting for the small possible disorder due to the randomised starting conditions. Specifically, the parameters of interest to us are the ones defined in equation (1.3) of the cited paper, namely:

\begin{equation}\label{eq:bo_par}
Q_l = \left(\frac{4\pi}{2l+1}\sum_{m=-l}^l|\overline{Q}_{lm}|^2\right)^{1/2}
\end{equation}

for any integer angular momentum channel $l$. However, at a difference with the original definition, here we slightly alter the way $\overline{Q}_{lm}$ is defined in order to include a smooth cutoff that only evaluates the local environment:

\begin{equation}
\overline{Q}_{lm} = \langle Q_{lm}(\vec{\mathbf{r}})\rangle = \frac{\langle S(\vec{\mathbf{r}})Y_{lm}(\theta(\vec{\mathbf{r}}), \phi(\vec{\mathbf{r}}))\rangle}{\langle S(\vec{\mathbf{r}})\rangle}
\end{equation}

with a sigmoid weighing function

\begin{equation}
S(\vec{\mathbf{r}}) = \frac{1}{2}\left[\frac{r_0-|\vec{\mathbf{r}}|}{\delta}\left[\left(\frac{r_0-|\vec{\mathbf{r}}|}{\delta}\right)^2+1\right]^{-1/2}+1\right]
\end{equation}

where we adopted $r_0 = 2 \angstrom$ and $\delta = 0.05 \angstrom$. In this case, equation \ref{eq:bo_par} was computed averaging over all bonds between the muon and the rest of the atoms in the unit cell (reduced to their closest periodic copy) for angular momentum channels ranging from $l=1$ to $l=6$, included. Therefore, the resulting gene had a length of six, which gave us a 7-dimensional array overall for each structure once the energy was included: $[Q_1,... , Q_6, E]$. Higher values of $l$ would increase sensitivity to small changes in the shape of the surrounding environment at the expense of a higher computational cost, which was deemed not necessary in this case. Therefore, this choice of genes is meant to highlight which output structures are similar to each other in both energy and local environment experienced by the muon.

\section{Results }\label{results}

\subsection{Hierarchical Clustering}\label{hierarchical clustering}

\subsubsection*{Silicon, Diamond and Germanium}

\begin{figure}[htp]
  \begin{center}
  \subfloat[Silicon]{\label{fig:Si_hier}\includegraphics[scale=0.55]{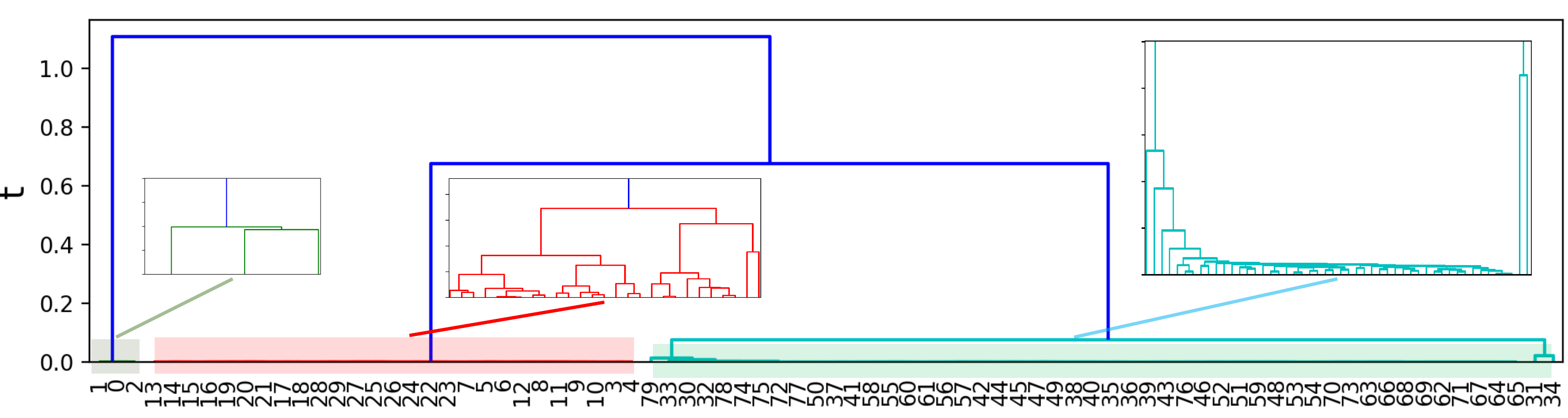}}\\ 
  \subfloat[Diamond] {\label{fig:Diam_hier}\includegraphics[scale=0.57]{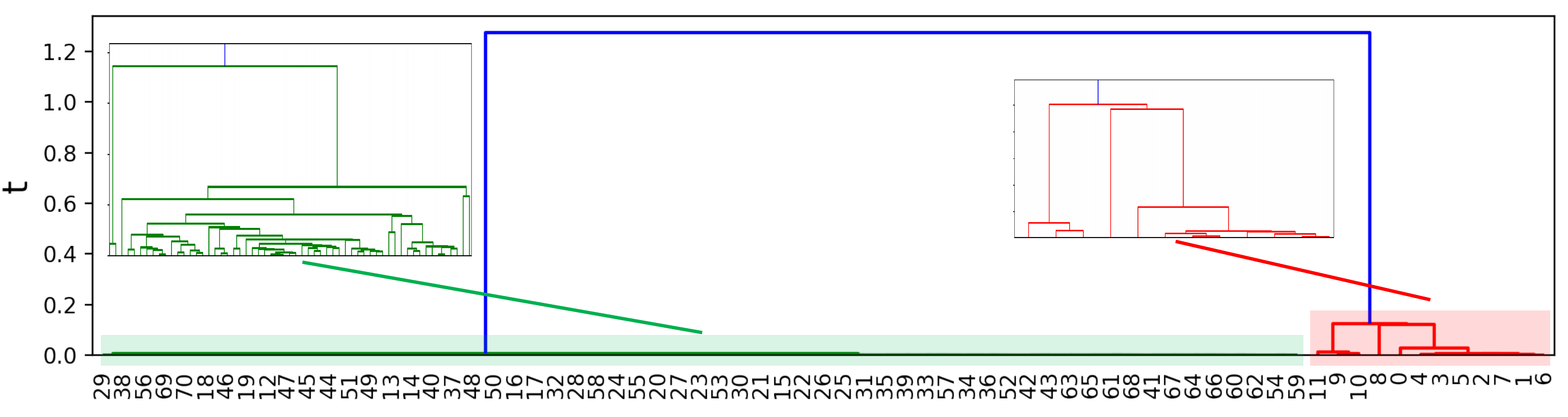}}\\ 
  \subfloat[Germanium]{\label{fig:Ge_hier}\includegraphics[scale=0.57]{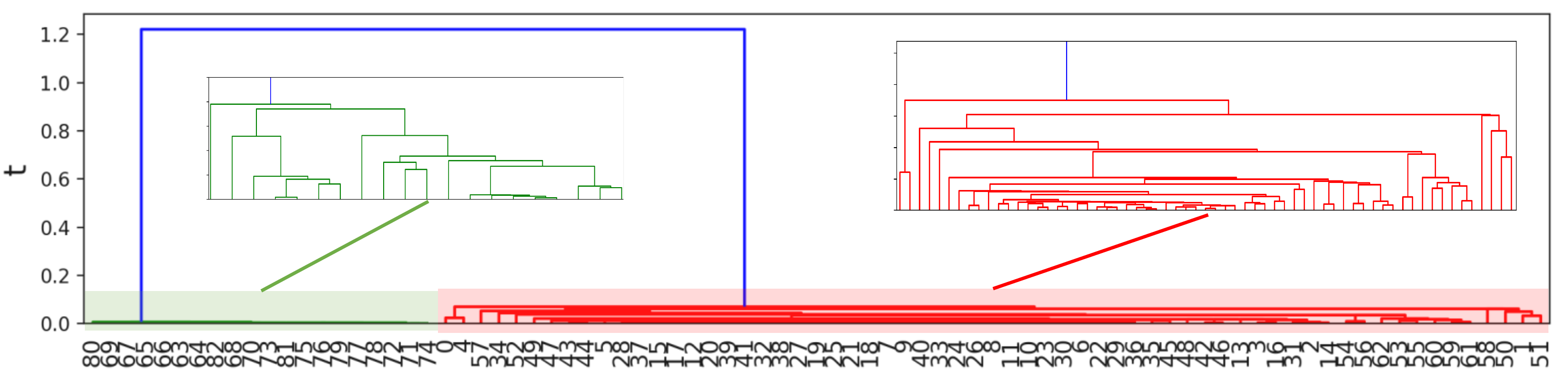}}
   \end{center}
  \caption{(\ref{fig:hier}) shows the hierarchical clustering trees for silicon, diamond and germanium, with colours applied for truncation at $t=0.2$ in the y axis. The blue lines above $t=0.2$ indicates the clusters into which the structures can be classified. The structures that resulted from the filtering process are labeled and placed along the x axis in accordance with their relative energies. The indexing starts with 0 for the lowest energy structure, but the values of these indexes are not correlated with their positions in the x axis: the structures with the lowest energies are closer to x=0.}
 \label{fig:hier}
\end{figure}

Figure (\ref{fig:hier}) shows the dendrogram results of hierarchical clustering for the silicon, diamond and germanium supercells.  The labels on the x axis correspond to all the structures that resulted from the filtering process, which have been sorted in accordance with their relative energies. On the y axis, it can be seen that the clustering starts at very small values of $t$ and, at $t\geq 0.1$, the systems can be clearly divided into 3 clusters for silicon and 2 clusters for diamond and germanium. To analyse the clustering behaviour at values of $t\leq 0.1$, we have zoomed in the regions in the x axis highlighted in grey, red and green rectangles: the branches of the dendrogram included in these rectangles are shown as insets in Figures (\ref{fig:Si_hier}), (\ref{fig:Diam_hier}) and (\ref{fig:Ge_hier}).\\

The low threshold for $t$ for cluster definition indicates that each of the clusters is composed of structures that look similar to each other in parameter space. The fact that the optimizations have converged, from  wildly different initial structures, to large sets with very small internal variability suggests that these likely correspond to muon stopping sites.  We will revisit this in the results of k-means clustering.\\

In the insets of figure (\ref{fig:Si_hier}) show the detailed structures of the clusters in silicon.  In the grey rectangle on the left hand side we can see a very well-defined cluster formed by three structures that is defined to a level of tolerance of $t\approx 0.002$. This cluster consists of the structures with indices 0, 1 and 2, which are also the three most stable structures and thus are likely to be defining a candidate stopping site. The next set of 27 structures highlighted in the red rectangle have higher energies.  However, their distribution in the parameter space is such that they are also defining a cluster to a level of tolerance of $t\approx 0.002$.  Finally, the 50 structures highlighted in green define a cluster to a level of tolerance of $t\approx 0.02$, but most of the structures in this cluster are already clustered at $t \approx 0.003$.  This distribution of structures indicates that some of the structures in the green cluster -the ones near the sides of the green rectangle- are slightly more 'dispersed' in the parameter space than the structures that compose the other two clusters.\\

Figures (\ref{fig:Diam_hier}) and (\ref{fig:Ge_hier}) show the dendrogram for the diamond and germanium  supercells. It shows 71 and 83 structures that at $t\geq 0.1$, can be clearly divided into 2 clusters, whose detailed branch structure for values of $t\leq 0.1$, are also shown in the insets of these figures. The clusters highlighted in green are defined for $t\approx 0.01$ and are composed of 59 structures in diamond and 20 structures in germanium, while the clusters highlighted in red are defined for $t\approx 0.1$ and are composed of 12 structures for diamond and 63 structures for germanium. Here the threshold values of $t$ for clustering differ in an order of magnitude, indicating a clear distinction between the structures of these clusters in the parameter space.  This distinction is clearly indicated in the results of k-means clustering.\\

\subsubsection*{Lithium Fluoride} \label{LiF-hierarchical}
 
\begin{figure}[htb]
  \begin{center}
 {\includegraphics[scale=0.57]{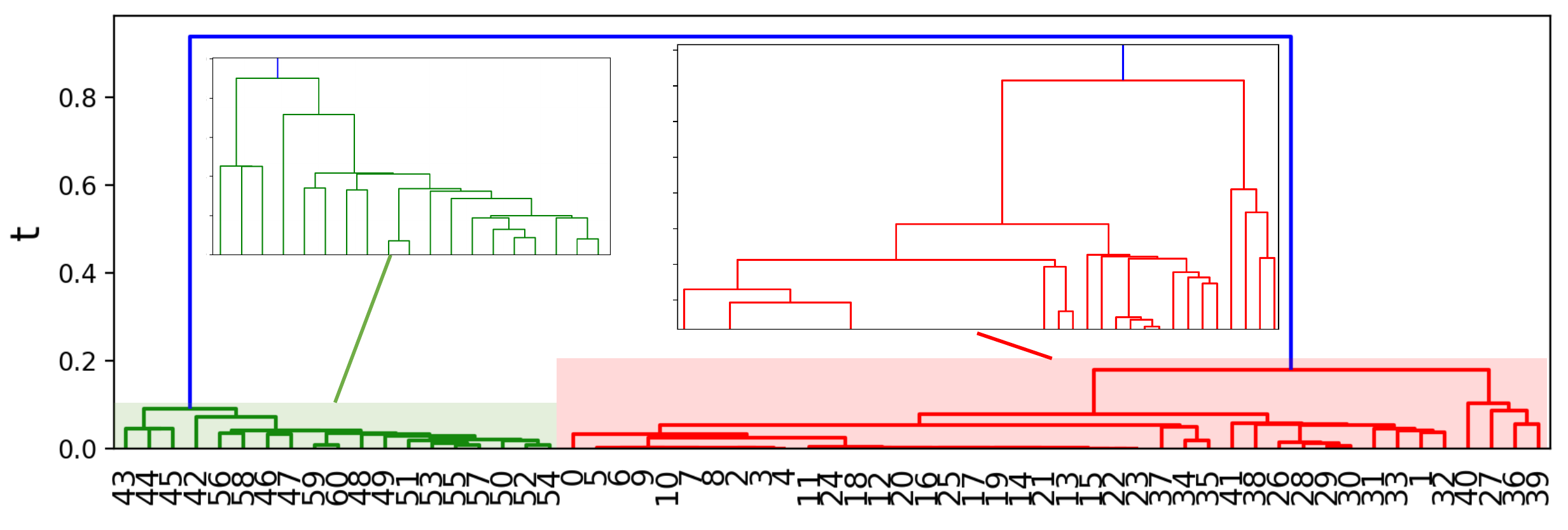}}
   \end{center}
  \caption{(\ref{fig:hier-LiF}) shows the hierarchical clustering tree for lithium fluoride, with colours applied for truncation at $t=0.2$ in the y axis. The blue lines above $t=0.2$ indicates the clusters into which the structures can be classified. The structures that resulted from the filtering process are labeled and placed along the x axis in accordance with their relative energies. The indexing starts with 0 for the lowest energy structure.}
 \label{fig:hier-LiF}
\end{figure}

 Figure (\ref{fig:hier-LiF}) shows the dendrogram for the lithium fluoride supercell. It shows 61 structures that at $t\geq 0.2$ can be separated into 2 clusters.  The insets in the figure show the detailed branch structure of these clusters. The cluster highlighted in  green is defined for $t\approx 0.1$ and is composed of 19 structures, while the cluster highlighted in red is defined for $t\approx 0.2$ and is composed of 42 structures.\\
 
\subsection{K-means Clustering} \label{k-means clustering}

\subsubsection*{Silicon, Diamond and Germanium}
 
\begin{figure}[htp]
  \begin{center}
  \subfloat[Silicon]{ \label{fig:Si_k} \includegraphics[scale=0.62]{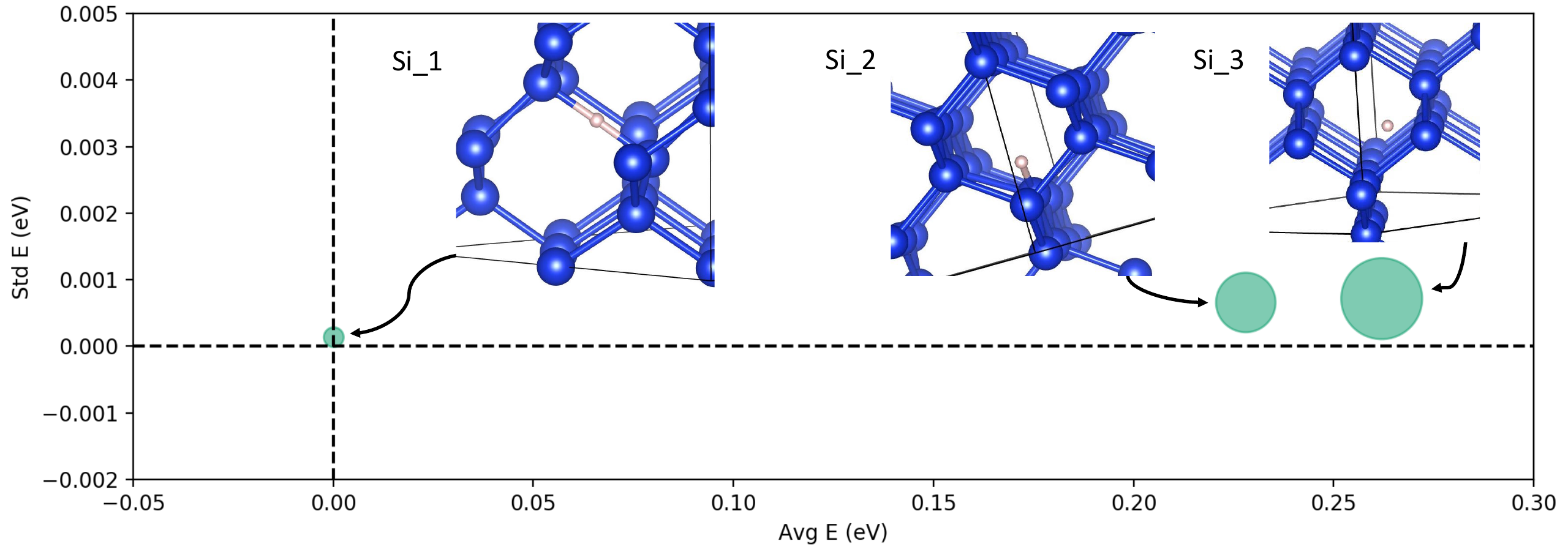}}\\ 
  \subfloat[Diamond]{ \label{fig:Diam_k} \includegraphics[scale=0.64]{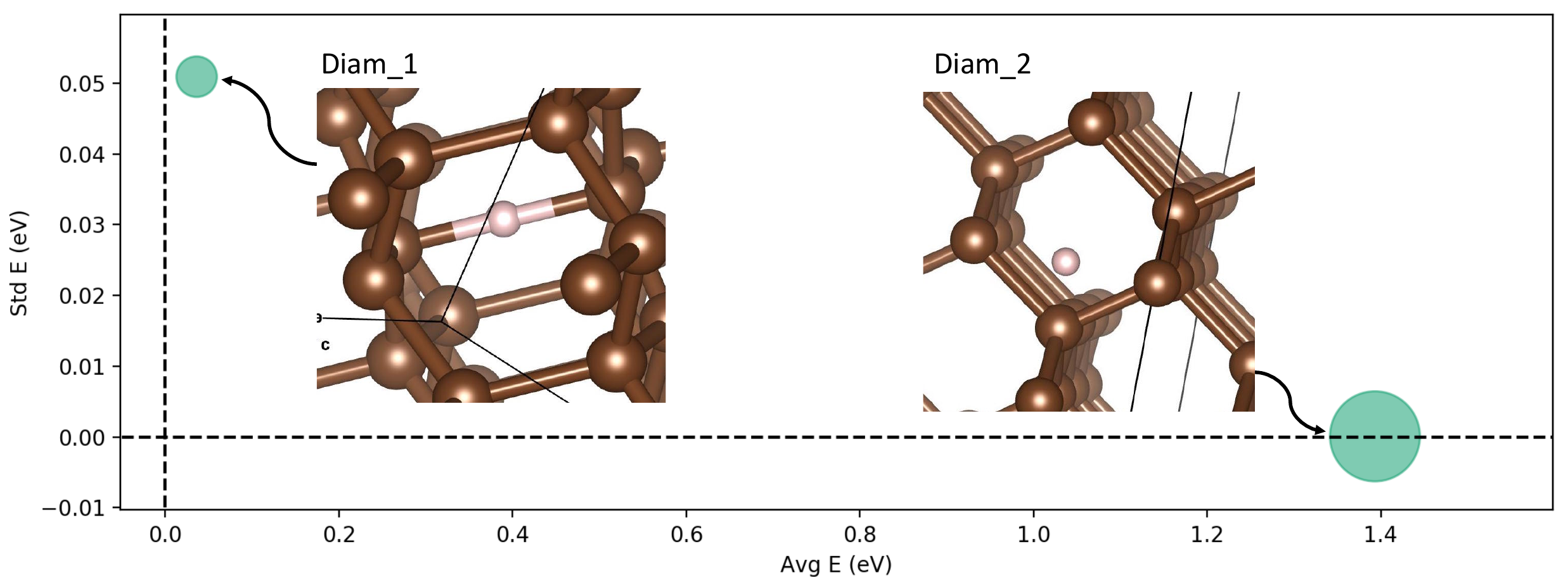}}\\ 
  \subfloat[Germanium]{\label{fig:Ge_k} \includegraphics[scale=0.61]{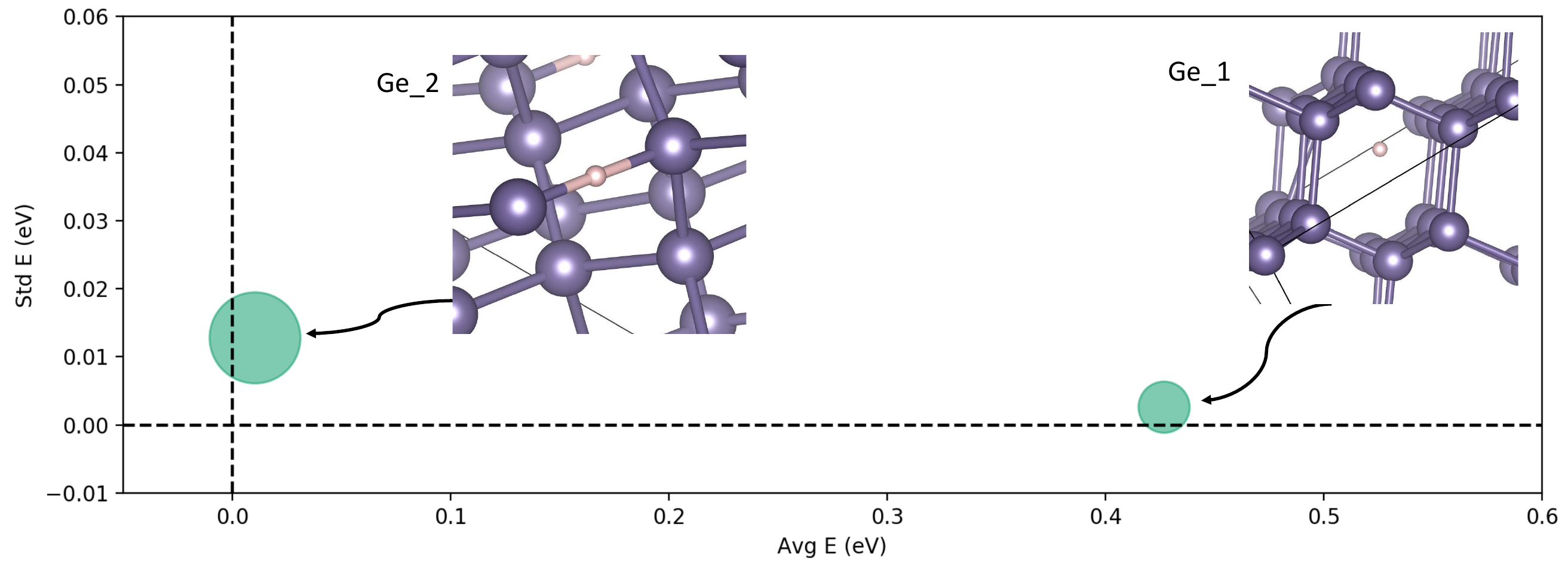}}
   \end{center}
  \caption{Figure (\ref{fig:k}) shows circles representing the clusters obtained via the k-means clustering method. The crystalline structures correspond to the most stable structure in each of the clusters.  The diameter of each circle represents the number of structures contained in each cluster.  The $x$ coordinate of the centre of each one of the circles indicates the average energy of the corresponding cluster -relative to the lowest energy structure in the cluster-, while the $y$ coordinate of the centre indicates the standard deviation of the average energy of that cluster.}
 \label{fig:k}
\end{figure}

Figure (\ref{fig:Si_k}) shows the k-means clustering in Si with a guess of n = 3.  The three clusters are represented by circles whose diameter is proportional to the number of structures contained in each cluster.  On the $x$ axis we indicate the average total energy of the structures belonging to each cluster relative to the lowest energy found in the system.  On the $y$ axis we indicate the standard deviation of the energy in each of the clusters.  Relatively small values for the standard deviations indicate consistent clusters, which are more likely to represent physical local energy minima. \\
 
\begin{figure}[htp]
  \begin{center}
   {\includegraphics[scale=0.25]{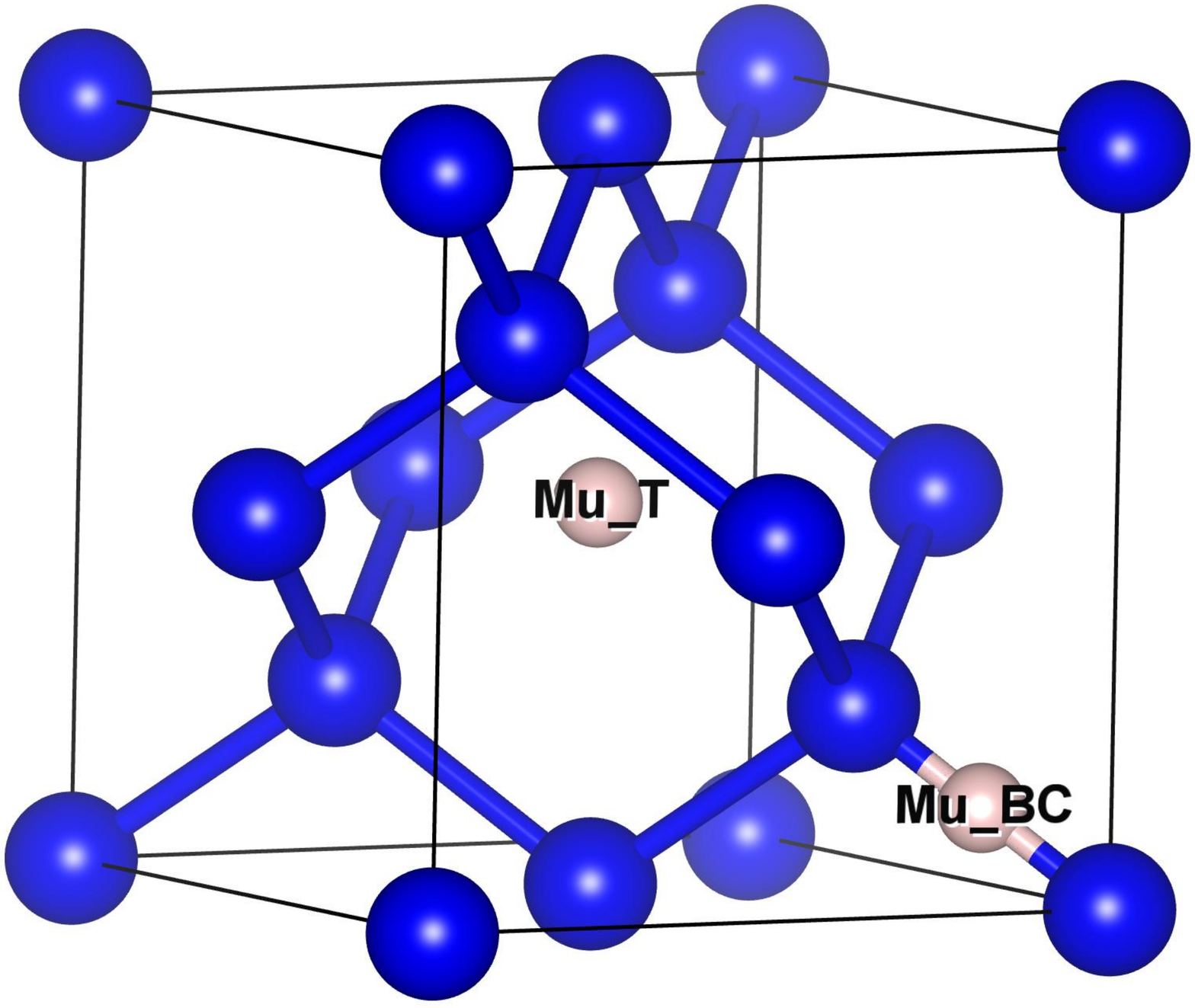}} 
    \end{center}
  \caption{ Tetragonal Mu$_{T}$ and bond-centred Mu$_{BC}$ sites in a generic conventional unit cell of a material with a cubic structure like Si, Diamond or Ge.  The space group is Fd3m and the only difference between these structures is that they have different lattice parameters.}
  \label{fig:Si-T-BC}
\end{figure}

Clusters \textsf{Si\textunderscore2} and \textsf{Si\textunderscore3} in figure (\ref{fig:Si_k}) contain structures where the muonium is within the tetrahedral/isotropic site (Mu$_{T}$) of the cubic cell structure type of Si, but with an important distinction. On one hand, cluster \textsf{Si\textunderscore2} is composed by the 27 structures showed in the red rectangle in figure (\ref{fig:Si_hier}). In these structures, the muon is bonded to one of the four Si atoms that define the tetrahedral site and is, therefore, away from the centre of the tetrahedron.  On the other hand, cluster \textsf{Si\textunderscore3}, composed of the 50 structures highlighted in green in figure (\ref{fig:Si_hier}), has its structures slightly displaced from the centre of the tetrahedron, but not bonded to any of the Si at the edges of the tetrahedral site.  According to these results, the absolute minimum for the muon stopping site in the tetragonal site of Si is not in the tetrahedral centre, but in a spherical shell surrounding a very low potential hill that is located at the tetrahedron centre.  This hill is low enough that the quantum effects associated with the muon would likely lead to a delocalisation of the muon that would stabilise the site and avoid symmetry breaking. In Appendix \ref{Tetragonal-site-Si} there is a more detailed discussion of these two alternative tetragonal sites for the muonium in Si. Regarding cluster \textsf{Si\textunderscore1} in figure (\ref{fig:Si_k}), it contains the 3 structures highlighted in grey in figure (\ref{fig:Si_hier}).  These structures are clustered around the axially symmetric bond-centred site (Mu$_{BC}$). The regions of a generic cubic cell structure corresponding to these muonium stopping sites, Mu$_{T}$ and Mu$_{BC}$, for isotropic and axially symmetric paramagnetic muoniums in Si, Diamond and Ge, are schematically shown in Figure (\ref{fig:Si-T-BC}). \\

Figures (\ref{fig:Diam_k}) and (\ref{fig:Ge_k}) show the k-means clusters obtained for diamond and germanium with a guess of n = 2. Clusters \textsf{Diam\textunderscore1} and \textsf{Ge\textunderscore2} contain 12 and 63 structures clustered around the axially symmetric bond-centred site (Mu$_{BC}$), and  clusters \textsf{Diam\textunderscore2} and \textsf{Ge\textunderscore1} contain 59 and 20 structures which have the muonium located near the centre of the tetrahedron and not bonded to any of the carbon or germanium atoms forming the tetrahedral site.  For diamond there is a difference of $\approx$1.4 eV between the relative average energies of the two clusters and of $\approx$0.05 eV between their corresponding standard deviations, with cluster \textsf{Diam\textunderscore1} being the most disperse cluster in the parameter space.  Regarding germanium, cluster \textsf{Ge\textunderscore2} has an average energy $\approx$0.41 eV larger than that of \textsf{Ge\textunderscore1} cluster.  The standard deviation of both \textsf{Ge\textunderscore1} and \textsf{Ge\textunderscore2} is below 0.02 eV, indicating a low dispersion for both clusters in the parameter space.\\

\subsubsection*{Lithium Fluoride} \label{LiF-k}
 
 \begin{figure}[htp]
  \begin{center}
 {\includegraphics[scale=0.62]{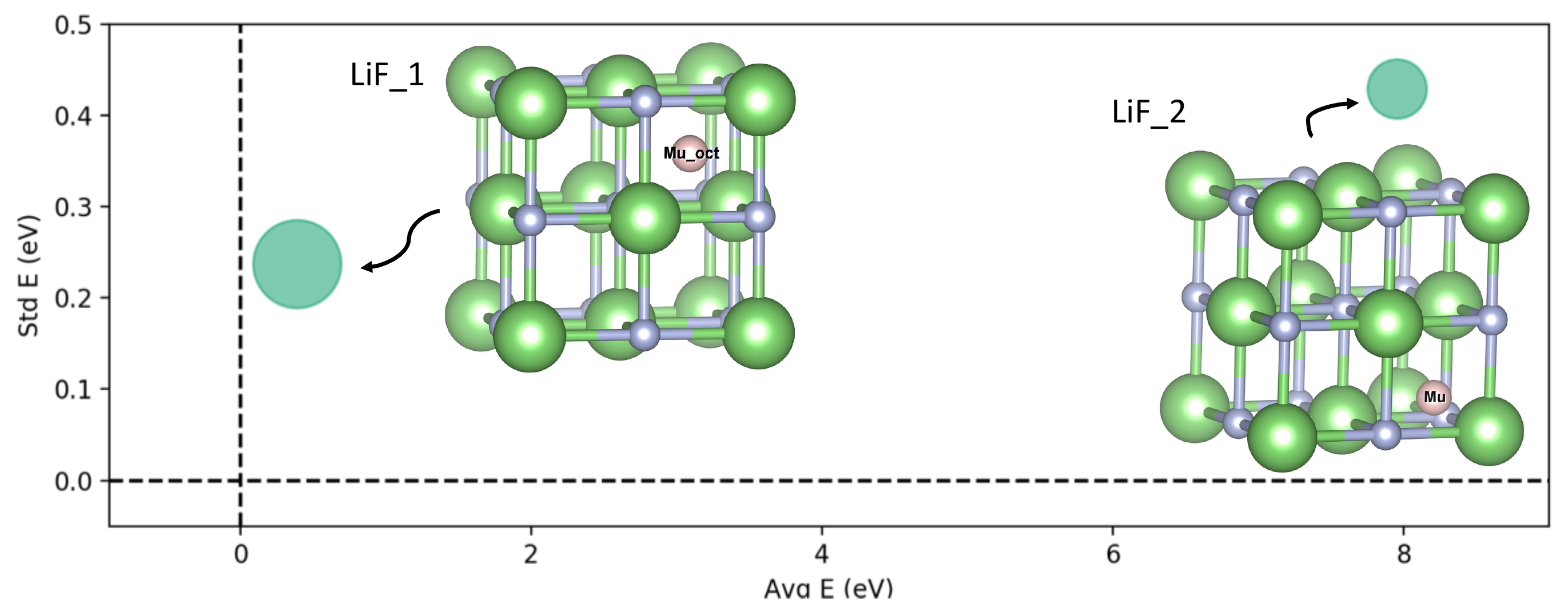}}\\ 
   \end{center}
  \caption{Figure (\ref{fig:k-LiF}) show  the two clusters obtained via the k-means clustering method in LiF. The $x$ coordinate of the centre of each one of the circles indicates the average energy of the corresponding cluster, while the $y$ coordinate indicates the standard deviation of the average energy of that cluster.  Clearly, cluster \textsf{LiF\textunderscore2} is too high in energy to be representing  a physically meaningful stopping site.}
 \label{fig:k-LiF}
\end{figure}
 
Figure (\ref{fig:k-LiF}) show the k-means clusters obtained for lithium fluoride with a guess of n = 2. Clusters \textsf{LiF\textunderscore1} and \textsf{LiF\textunderscore2} contain 42 and 19 structures respectively.  Cluster \textsf{LiF\textunderscore1} has an average energy of $\approx$ 0.38 eV and a standard deviation of $\approx$0.23 eV, and its structures are clustered around the octahedral interstitial site \textcolor{red}{(Mu$_{oct}$).}  \textcolor{red}{ The radial displacements of the first-neighbours to the Mu$_{oct}$ site, with respect to the unperturbed LiF cell, are of $\approx$0.02\AA~for the Fluorine atoms and of  $\approx$0.017\AA~for the Lithium atoms, which is in agreement with previous results \cite{moller2013quantum}. At these short distances the displacements of the neighbouring atoms are pointing outwards in the radial direction from the muon.}  On the other hand, cluster \textsf{LiF\textunderscore2} has an average energy of $\approx$8.0 eV and an standard deviation of $\approx$0.43 eV: all the structures in this cluster are much more dispersed in the space and have a much larger energy.  The image shown in figure 
(\ref{fig:k-LiF}) indicates the structure with the minimum energy in the cluster, which has lower symmetry than the other. In general, due to the high energies involved, it seems reasonable to consider this cluster as non physical, or at least not experimentally relevant.\\
 
\section{Analysis}\label{analysis}

Using only ab initio simulations and data analysis techniques we were able to identify a small number of candidate stopping sites for muonium in crystalline silicon, germanium, diamond and lithium fluoride. Each site is represented by a cluster of size ranging from 3 to 60 optimised structures, identified by energetic and geometric similarities. The full list of their properties is listed in Table \ref{tab2}.  The low values of the standard deviation of energy among these clusters reinforce the case that they represent indeed small fluctuations around a single energy minimum, and are not just flukes. If, for example, two non-equivalent minima were grouped under the same cluster due to accidental geometric similarities, we would expect a much higher dispersion of the energy values.\\

The predicted sites, that can be seen in figures (\ref{fig:Si_k}), (\ref{fig:Diam_k}), (\ref{fig:Ge_k}) and (\ref{fig:k-LiF}), closely match what is known from the literature about muonium defects in diamond, silicon, germanium and lithium fluoride crystals from both experiments and theoretical calculations \cite{Patterson:1988aa, luchsinger1997gradient, moller2013quantum}, which have identified a bond-centred (Mu$_{BC}$) and a tetragonal(Mu$_{T}$) stopping site diamond, silicon and germanium, and an octahedral site \textcolor{red}{(Mu$_{oct}$)} in lithium fluoride.
\begin{center}
\begin{table} 
\begin{tabular}{ | c | c | c | c | }
\hline
  Cluster & Rel. Aver. Energy (eV) & Structures in Cluster & Standard Dev. (eV)\\
\hline
\textsf{Si\textunderscore1}  & 0 & 3 & 0.0002 \\
\textsf{Si\textunderscore2}  & 0.23 & 27 & 0.0012 \\
\textsf{Si\textunderscore3}  & 0.26 & 50 & 0.0015 \\
\textsf{Diam\textunderscore1} & 0.02 & 12 & 0.051 \\
\textsf{Diam\textunderscore2} & 1.4 & 59 & 0.01 \\
\textsf{Ge\textunderscore1} & 0.42 & 20 & 0.005 \\
\textsf{Ge\textunderscore2} & 0.01 & 60 & 0.015 \\
\textsf{LiF\textunderscore1} & 0.38 & 42 & 0.23 \\
\textsf{LiF\textunderscore2} & 0.79 & 19 & 0.43 \\
\hline  
\end{tabular}
\caption{ Main properties of all of the identified clusters in Si, Diamond,  Ge and LiF.} 
\label{tab2}
\end{table}
\end{center}

Regarding the sites in silicon, diamond and germanium, we know from the literature\cite{Patterson:1988aa, 0953-8984-16-40-017} that both the (Mu$_{BC}$) and (Mu$_{T}$) sites were experimentally observed. The data from Table\ref{tab2} however suggests that the tetrahedral sites (represented by clusters \textsf{Si\_2} and \textsf{Si\_3}, \textsf{Diam\_2}, and \textsf{Ge\_2}) all have higher formation energies than the bond centred ones. The lowest energy difference between tetrahedral and bond centred sites, $\Delta E$, is 0.23 eV. If we assume that the system is in thermal equilibrium, this value of $\Delta E$ implies that a temperature of more than 2000 K is needed to transition from one stopping site to the other.  Hence, the order of magnitude alone seems to completely refute the possibility that the tetrahedral site could be observed in thermal equilibrium at low temperatures. The prediction that the Mu$_{BC}$ site is lower in energy than the Mu$_T$ one  is also in qualitative agreement with previous theoretical results \cite{luchsinger1997gradient}. This reinforces the commonly held view that during an experiment muons do not have the time to effectively relax to their equilibrium state and can therefore occupy metastable sites \cite{0953-8984-16-40-017}.\\

Furthermore, our calculations support the prediction of delocalisation of the muon in the Mu$_{T}$ site of Si that was advanced in previous studies \cite{Miyake:1998aa, PhysRevB.60.13534}. In order to clarify this point, we carried out further calculations on many configurations surrounding the site to plot the local energy landscape, reported in Appendix \ref{Tetragonal-site-Si}. We observed for the Mu$_{T}$ site in Si a flat potential with a slight maximum in the centre and a minimum approximately distributed in a radial shell surrounding it; this is the most likely reason for the observation of two clusters corresponding to that site, \textsf{Si\_2} and \textsf{Si\_3}, with different positions for the muonium itself with respect to the centre of the tetrahedra. Experimentally, the delocalisation of the muon in the tetragonal site was first proposed by Holzschuh \textit{et. al.} \cite{holzschuh1983direct}. In this model, the muon hops between different sites in the tetragonal region, which helped to explain the anomalous temperature dependence of the isotropic component of the hyperfine coupling constant in Si.\\

Regarding the stopping sites in lithium fluoride, our methodology predicted the octahedral stopping site \textcolor{red}{(Mu$_{oct}$)} for muonium, which is the site that has been both experimentally and theoretically predicted \cite{moller2013quantum, LiF}. The other local minimum we found has an average energy too large to be a physically meaningful muonium stopping site.

\section{Conclusions}

We have proposed a purely theoretical method to predict muon stopping sites in crystalline materials. The method is based on a combination of ab initio random structure searching and machine learning, and it has successfully predicted the Mu$_{T}$ and Mu$_{BC}$  stopping sites of muonium in Si, Diamond and Ge, and the octahedral stopping site \textcolor{red}{(Mu$_{oct}$)} in lithium fluoride, purely from first principles. The process is easily reproducible and requires little human input to analyse dozens or even hundreds of structures. Soprano, a Python library containing the tools used for this analysis as well as many others designed for different systems, has been released publicly \cite{soprano} and will be fully documented in a future work.\\

\section{Acknowledgements}

The authors would like to thank Barbara Montanari, from the Scientific Computing Department at RAL, and Francis Pratt and Stephen Cottrell, from the Muons Group at ISIS,  for the useful discussions.  The authors are also grateful for the computational support provided by: (a) STFC Scientific Computing Department's SCARF cluster; (b) the UK national high performance computing service, ARCHER, for which access was obtained via the UKCP consortium and funded by EPSRC grant ref EP/K013564/1; and (c) the UK Materials and Molecular Modelling Hub for computational resources, which is partially funded by EPSRC (EP/P020194/1). Funding for this work was provided by STFC-ISIS muon source and by the CCP for NMR Crystallography, funded by EPSRC grants EP/J010510/1 and EP/M022501/1.

\appendix

\section{The \texttt{phylogen} module in Soprano}\label{app:phylogen}

Soprano's \texttt{phylogen} module contains classes and functions for clustering similarity analysis of large populations ofanalyse structures. It owes its name to the analogy between this approach and the construction of phylogenetic maps or trees connecting living species based to the similarity of their DNA. Similarly, in this module, the user can pick a number of `genes' by which the structures should be characterised, and these genes will be chained together into a single array that uniquely defines each structure. These arrays will then constitute the points whose similarity relationships are evaluated through clustering.\newline
In Soprano, a gene can be any scalar or vectorial property of a structure. For example, the energy of a structure can be a gene of only one element; its lattice parameters $a,b,c$ a gene of three; and so on. When performing a clustering operation the user can also pick a weight for each gene, thus determining the relative importance between them, and a normalising interval. By default, all genes are weighted equally, and they're normalised to $[0, 1]$ over the entire collection being analysed. So, for example, if we indicate the raw value for element $j$ of gene $i$ as $x_i^j$ in an analysis using $g$ genes each with $n_i$ elements we'll have that the `DNA' array for a given structure is defined as:

\begin{equation}
[X_1^1, X_1^2, ..., X_1^{n_1}, X_2^1, ..., X_2^{n_2}, ..., X_g^{n_g}]
\end{equation}

The normalised and weighted values $X_i^j$ are defined as:

\begin{equation}
X_i^j = \frac{w_i}{\sqrt{n_i}}\frac{[x_i^j-\min(x_i^j)]}{[\max(x_i^j)-\min(x_i^j)]}
\end{equation}

where $w_i$ is the user-defined weight, minima and maxima refer to the whole collection, and the square root term has the purpose of normalising for size, to prevent longer genes from dominating the classification.

\section{Tetragonal/isotropic Mu$_{T}$ site in Si}\label{Tetragonal-site-Si}

\begin{figure}[htp]
  \begin{center}
   \includegraphics[scale=0.4]{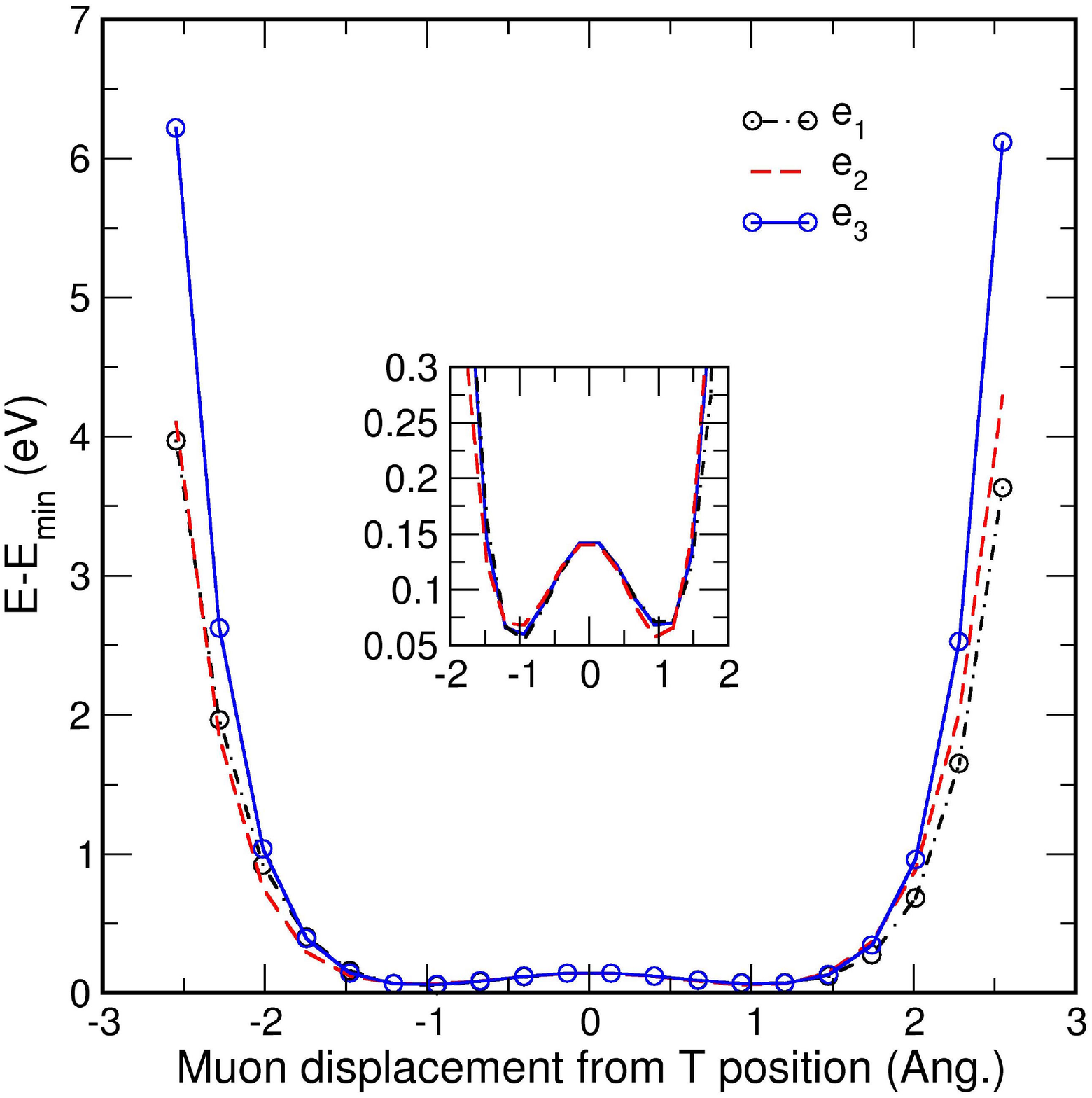}
    \end{center}
  \caption{DFT total energies calculated at equally spaced points along the three eigenvectors corresponding to the calculated negative frequency for Mu$_{T}$ }
  \label{fig:BO_surfaces}
\end{figure}

We investigated the nature of the tetrahedral site Mu$_{T}$ by performing CASTEP phonon calculations at the $\Gamma$ point on a Si supercell with the muon perfectly located in the tetragonal site.  This muonated Si supercell has 195 vibrational modes, and we assumed that the modes $\omega_{i}$ associated with the muon are mutually perpendicular and are decoupled from all the other modes as a consequence of its much smaller mass.   However, we found that all three the main vibrational frequencies for Mu$_{T}$ are negative, which indicates that the associated mode is imaginary, i.e.: the muon in the tetragonal position is in a maximum of its corresponding Born-Oppenheimer (BO) potential.  To estimate the size of this maximum, we took the calculated eigenvectors corresponding to those negative frequencies and displaced the muon along them, in positive and negative directions, and then calculated at equally spaced points the total DFT energy.  Figure (\ref{fig:BO_surfaces}) shows the results of these calculations. The shape depends only minimally from the specific eigenvector chosen, which suggests that the potential is close to a radially symmetric quartic with a minimum shell around $r=1\,\angstrom$.

This maximum for the BO potential in the Tetragonal site of muonium has been observed previously using DFT calculations \cite{Miyake:1998aa, luchsinger1997gradient}, and there is experimental and theoretical evidence that the muon in the T site of Silicon is delocalised \cite{Miyake:1998aa, holzschuh1983direct}.  To the best of the author's knowledge, there is only one theoretical paper that reports a minimum of the BO potential in the tetragonal site \cite{PhysRevB.60.13534}, but this result could not be confirmed by other DFT calculations.  Furthermore, we conducted computational tests to rule out that the maximum in the BO potential is an artificial result arising from our calculations.  We performed full geometrical relaxations and phonon calculations at the $\Gamma$ point, for Si with Mu in the tetragonal site using the LDA and GGA functionals, and obtained the results indicated in Table {\ref{table:LDA-GGA}}.

\begin{table} [ht]
\centering
\begin{tabular}{| c | c | c | c | c |}
\hline
Si                   			& Exp.  	& DFT-LDA  	& DFT-GGA  & DFT-GGA (a from LDA) \\ \hline
 a $(\angstrom)$  		&    5.43    &  5.40     	&    5.47       &   5.40  \\  
f$_{1}$ $(cm^{-1})$		&  	N/A	& -182.13		 &    -410.0   &   -409.9 \\
f$_{2}$ $(cm^{-1})$		&	N/A	  & -182.1		 &     -409.9  &   -409.9 \\
f$_{3}$ $(cm^{-1})$		&	N/A	  & -182.1		 &     -409.9  &    -409.8 \\ \hline
 \end{tabular} 
 \caption{ Lattice parameters and negative (imaginary) phonon frequencies for muonium in the tetragonal site of Si.}
 \label{table:LDA-GGA}
\end{table}

\bibliographystyle{unsrt}

\end{document}